\documentclass[conference]{IEEEtran}
\IEEEoverridecommandlockouts
\usepackage{etex,etoolbox}
\usepackage{amsthm,amssymb}
\usepackage{thmtools}
\usepackage{environ}

\usepackage{subfig}
\usepackage[numbers]{natbib}
\usepackage{graphicx}

\usepackage{mathtools}
\usepackage{commath}
\usepackage{dsfont}
\usepackage{booktabs} 
\usepackage{array} 
\usepackage{paralist} 
\usepackage{verbatim} 
\usepackage{algorithm}
\usepackage{multirow}
\usepackage{authblk}

\usepackage{algpseudocode}
\let\oldReturn\Return
\renewcommand{\Return}{\State\oldReturn}
\usepackage{tikz}

\usepackage{xcolor}

\makeatletter
\providecommand{\@fourthoffour}[4]{#4}

\newtheorem{thm}{Theorem}
\newtheorem{pro}{Proposition}
\newtheorem{definition}{Definition}
\def\thanks#1{\protected@xdef\@thanks{\@thanks
        \protect\footnotetext{#1}}}
\makeatother

\begin{document}
%
\title{Windowed total variation denoising and noise variance monitoring}


\author[1,3]{Zhanhao~Liu}
\author[3]{Marion~Perrodin}
\author[2]{Thomas~Chambrion}
\author[1]{Radu~S.Stoica}

\affil[1]{Université de Lorraine, CNRS, Inria, IECL, F-54000 Nancy, France}
\affil[2]{Institut de Mathématiques de Bourgogne, UMR 5584, CNRS, UBFC, F-21000 Dijon, France}
\affil[3]{Saint-Gobain Research Paris, 39 Quai Lucien Lefranc, F-93300 Aubervilliers, France}


%


\maketitle

\begin{abstract}
We proposed a real time Total-Variation denosing method with an automatic choice of hyper-parameter $\lambda$, and the good performance of this method provides a large application field. In this article, we adapt the developed method to the non stationary signal in using the sliding window, and propose a noise variance monitoring method. The simulated results show that our proposition follows well the variation of noise variance.
\end{abstract}


%
\IEEEpeerreviewmaketitle

\section{Introduction}

The signal $y = (y_1, \cdots, y_n) \in \mathbb{R}^n$ collected by the sensor can be modeled as $y = u + \epsilon$: a random noise $\epsilon$ is added into the useful physical quantity $u$ with $\mathbb{E}(\epsilon) = 0$ and $\mathbb{V}(\epsilon) = \sigma^2$.

We aim to recover the unknown vector $u=(u_1, \cdots, u_n)\in \mathbb{R}^n$ from the noisy sample vector $y = (y_1, \cdots, y_n)$ with $y_i$ the sample at time $t_i$ by minimizing the Total Variation (TV) restoration functional:
\begin{align}
    F(u, y, \tau, \lambda) = \sum_{i = 1}^n \tau_i(y_i-u_i)^2  + \lambda  \sum_{i=1}^{n-1}|u_i-u_{i-1}|
    \label{equ:tv}
\end{align}
with the sampling period vector $\tau = (\tau_1, \cdots, \tau_n)$ where $\tau_i = t_{i}-t_{i-1}$ for $i = 2,..., n$ and $\tau_1 = \tau_2$.
The restored signal is given by: 
\begin{equation}
    u^*(\lambda)=(u^*_1(\lambda), \cdots, u^*_n(\lambda)) = \arg\min F(u, y,\tau, \lambda)
    \label{equ:estimation}
\end{equation}

 Total Variation based restoration method is first proposed in \cite{rudin1992nonlinear}.  The authors in \cite{chambolle1997image} show an one-to-one correspondence between the noise's variance $\sigma^2$ and $\lambda$ under the hypothesis of constant $\sigma^2$. 


Motivated by the local influence of new sample to the actual restoration, we propose a new online TV-restoration algorithm with an automatic choice of $\lambda$ for a stationary 1D signal in \cite{Liu} following the work of \cite{Tibshirani} and \cite{pollak2005nonlinear}. The simulations show that our proposition of $\lambda$ has a similar performance as the existing methods (SURE \cite{stein1981estimation} and cross-validation). Further more, our method is appropriate for real time applications, especially for monitoring a huge amount of sensors in the plants.

In this paper, we adapt our online TV restoration to non-stationary signals and propose a noise variance monitoring method. The key idea is to use our method on sliding windows, assuming a local-stationary hypothesis for the signal. This hypothesis is strong but commonly used already (e.g. Fourier analysis, wavelet transforms). In order to apply our method to non-stationary signals, two problems are to be solved:
\begin{itemize}
    \item Adaptation of the online method to the sliding window
    \item Choice of the length of window (noted $m$)
\end{itemize}
The first point ensures the efficiency of the real time estimation, while the second point guarantees the good performance of method. Indeed, the choice of $m$ needs to get a compromise between the local stationarity assumption and the performance of the restoration : the local influence of new sample vanishes inside a large window, but the local stationarity may not be respected. In this article, we present how we deal with the first point by supposing the influence of new sample stays inside the actual window. The choice of window size is the immediate perspective of this work.

The increasing of the ground noise is one of the index for the failures of the sensors or the production lines. This motivates us to build a real time application for detecting the variance behaviour by tracking the windowed restoration residuals $y-u^*(\lambda_{ours})$ based on the automatic determination of hyper-parameter $\lambda_{ours}$.

Here some approaches for the estimation of noise variance $\sigma^2$ are listed:
\begin{itemize}
    \item Most of the existing methods are based on the separation of signal and noise by digital filters \cite{shin2005block} or the restoration methods (i.e. wavelet transform) \cite{beheshti2005new} \cite{Hashemi} \cite{Han}. See more details in \cite{pyatykh2012image}. 
    \item  Median absolute deviation (\emph{MAD}) proposed by \cite{donoho1994ideal} is a heuristic method largely used in many applications, especially for the choice of threshold for wavelet restoration. $\sigma$ can be estimated by the median absolute deviation of the finest scale of wavelet coefficients $By$ of signal.
\end{itemize}

With respect to the previous cited method, our approach, based on the separation of signal and noise, proposes a procedure for simultaneously signal denoising and variance tracking, while being implementable as an online application. Thanks to its low time and space complexity, our proposition is well fitted to the cases with limited computation resource. 

The paper continues as it follows: in Section \ref{sec:dpvt} we will at first present our TV-based denosing method with an automatic choice of parameter. Then, in Section \ref{sec:adapt}  we adapt our online implementation to the sliding window. After that, in Section \ref{sec:app}, we will present an application of our algorithms: noise variance monitoring, and compare with an existing method. Finally, conclusions and perspectives are depicted.

%

\section{Automatic total variation denoising method}
\label{sec:dpvt}
In this section, we will present the automatic TV-denoising method proposed in \cite{Liu}. We aim to recover the unknown vector $u$ from the noisy samples $y$ by minimizing (\ref{equ:tv}). The restored signal is given by (\ref{equ:estimation}).

 Since we work on a finite sample of signal, the solution $u^*(\lambda)$ can be seen as piece-wise constant. We use the \emph{segment} representation \cite{pollak2005nonlinear} for the constant pieces: a set of index $\{j, j+1, \cdots, k\}$ of consecutive points whose restored value $u^*_j(\lambda) =\cdots = u^*_k(\lambda)$ is called a \emph{segment} if it can not be enlarged, which means if $u^*_{j-1}(\lambda) \neq u^*_j(\lambda)$ (or $j =1$) and $u^*_{k}(\lambda) \neq u^*_{k+1}(\lambda)$ (or $k =n$). The segments number of $u^*(\lambda)$ is noted as $K(\lambda)$. The following notations are introduced for the $j^{th}$ segment:
\begin{itemize}
    \item Index set $\mathcal{N}_j(\lambda) = \{i^{j}_1, \cdots, i^{j}_{n_j} \}$ with $n_j(\lambda) = \text{Card}(\mathcal{N}_j(\lambda))$, containing the point inside the $j^{th}$ segment.
    \item Segment level $v^*_j(\lambda) = u^*_i(\lambda), \forall i \in \mathcal{N}_j(\lambda)$ 
\end{itemize}

An equivalent representation of $u^*(\lambda)$ is provided by the couple $\{v^*(\lambda), \mathcal{N}(\lambda)\}$ with the set of \emph{segment levels} $v^*(\lambda) =(v^*_1(\lambda), \cdots, v^*_K(\lambda))$ and the \emph{cutting set} $\mathcal{N}(\lambda) = \{\mathcal{N}_1(\lambda), \cdots, \mathcal{N}_K(\lambda)\}$. By knowing the cutting set $\mathcal{N}(\lambda)$, let $s(v) = \{s_0(v), s_1(v), \cdots, s_K(v)\}$ with $s_i(v) = \text{sign}(v_{i+1}-v_i)$ for $i=1,\cdots,K-1$, $s_i(v) =0$ for $i\in \{0, K\}$ 
and $s^* = s(v^*(\lambda))$, the level of $j^{th}$ segment is given by:
\begin{equation}
    v^*_j(\lambda) =  \overline{y}^*_j + \frac{\lambda}{2\mathcal{T}_j} (s^*_{j} - s^*_{j-1})
    \label{equ:v_solution}
\end{equation}
with $j =1,\cdots,K(\lambda)$, the segment length $\mathcal{T}_j = \sum_{i\in \mathcal{N}_j(\lambda)}\tau_i$ and the mean value inside $ j^{th}$ segment $\overline{y}^*_j =  \frac{\sum_{i\in \mathcal{N}_j(\lambda)}\tau_i y_i}{\mathcal{T}_j}$. 


A different value of $\lambda$ may provide a solution with distinct cutting set: $\lambda = 0$ gives $u^* = y$, while $\lambda = \infty$ implies a parsimonious solution $u^* = \text{mean}(y)$. The authors in \cite{Tibshirani} and \cite{pollak2005nonlinear} show there exists a sequence $\Lambda = (\lambda_1, \cdots)$ such that for every $\lambda \in \Lambda$, two segments are merged together by moving $\lambda-\eta$ to $\lambda$ with $\eta \to 0$ and $\eta >0$. In \cite{Liu}, we propose a rapid algorithm to estimate the dynamic of the restoration in function of $\lambda$, and the dynamic is saved in $\Lambda^{\circ} = (\lambda^{\circ}_1, \lambda^{\circ}_2, \cdots, \lambda^{\circ}_{n-1})$ with $\lambda^{\circ}_i$ the value of $\lambda$ for which the points $i$ and $i+1$ are merged into the same segment. $\Lambda^{\circ}$ allows the computation of the cutting set and also the restoration (\ref{equ:v_solution}) for every $\lambda$.

Based on the variation of the extremums number (noted $g(\lambda)$) of the restoration $u^*(\lambda)$ in function of $\lambda$, an adaptive choice of hyper-parameter $\lambda$ is proposed. The simulations show that our estimation $\lambda_{ours}$ has a similar performance as the state of the art, all near the optimal choice $\lambda_{op} = \arg\min|u^*(\lambda) - u_{net}|^2$ with $u_{net}$ the original signal. The variation $\Delta g(\lambda) = g(\lambda-\eta) - g(\lambda)$ for every $\lambda \in \Lambda$ with $\eta \to 0$ and $\eta>0$ can be estimated simultaneously as the estimation of $\Lambda^\circ$.  

Besides, we analysed the local influence for $\Lambda^{\circ}$ by introducing a new sample at the end of the sequence: only a small part of $\Lambda^{\circ}$ will be changed. Based on this local property, we proposed an online algorithm (c.f. Algorithm 2 in \cite{Liu}) for updating the changed part of $\Lambda^{\circ}$, which is more efficient than the offline estimation. The locality of the TV-denoising method motives the adaption of sliding windows in order to deal with the non-stationary signal by supposing the local stationarity inside the window.


\section{Adaptation to sliding windows}
\label{sec:adapt}
In this section, we will present some theoretical elements about the local behaviour of the restoration and adapt the online algorithm to the sliding window.

Let's consider two successive sliding windows $ w_i^m = [i, i+m-1]$ and $w_{i+1}^m = [i+1, i+m]$. $w_{i+1}^m$ is indeed $w_{i}^m$ in which we take off the first point $(y_i, \tau_i)$ and add a new point $(y_{i+m}, \tau_{i+m})$. For the sake of simplicity, $F(u_{\{i, \cdots, i+m-1\}},y_{\{i, \cdots, i+m-1\}},\tau_{\{i, \cdots, i+m-1\}}, \lambda )$ is noted  $F^m_i(\lambda)$ for the window $w^m_i$. 

We note $u^* =(u^*_1, \cdots, u^*_m) = \{v^*, \mathcal{N}^*\} = \arg\min F^m_i$ the restoration of the window $w^m_i$ with a given $\lambda$ and $\hat{u}= (\hat{u}_1, \cdots, \hat{u}_m)= \{ \hat v, \hat{\mathcal{N}}\}= \arg\min F^m_{i+1}$ for $w^m_{i+1}$. For the simplify of the presentation, the elements concerning about $w^m_i$ will be noted with the symbol $*$, and that for $w^m_{i+1}$ will be noted with the hat symbol. For example, with a given $\lambda$, the segment length set of $u^*(\lambda)$ is $\mathcal{T}^* = \{\mathcal{T}^*_1, \cdots, \mathcal{T}^*_{\hat{K}}\}$ with $\mathcal{T}^*_j = \sum_{i\in \mathcal{N}^*_j} \tau_i$, and that of $\hat u(\lambda)$ is  $\hat{\mathcal{T}} = \{\hat{\mathcal{T}}_1, \cdots, \hat{\mathcal{T}}_{\hat{K}}\}$ with $\hat{\mathcal{T}}_j = \sum_{i\in \hat{\mathcal{N}}_j} \tau_i$.


\subsection{Influence of slide movement}
The update of the restored signal and $\Lambda^\circ$ need to consider the sliding of index from $w^m_i$ to $w^{m}_{i+1}$: a restored signal point is unchanged under the influence of slide movement means $\hat u_{i-1} = u^*_i$. In \cite{Liu}, we have shown the following theorem: 
\begin{thm}
\label{thm:diff_bloque}
    If there exists an index $j \in \{2, \cdots, K^*(\lambda)\}$ such that $\text{sign}(v^*_{j-1}(\lambda)- v^*_{j}(\lambda))= \text{sign}(v^*_{K^*(\lambda)}(\lambda) - y_{i+m})$, then the new restoration $\hat{u}$ satisfies $\hat u_{i-1}(\lambda) = u^*_i(\lambda)$ for all $i<i^{j,*}_1$.
\end{thm}
The diffusion from the ``new'' sample $y_{i+m}$ changes only the last part of the restoration of $w^i_m$ up to the junction of two segments whose sign of the variation $v^*_{j-1}(\lambda)- v^*_{j}(\lambda)$ corresponds to that of $v^*_{K^*}(\lambda) - y_{n+1}$. We can establish a similar theorem (c.f. Theorem \ref{thm:diff_bloque_head}) about the influence of the first sample $y_i$ for the sliding window $w^m_i$: the removing of $y_i$ changes only the first part of the restoration.

\begin{thm}
    If there exists an index $j \in \{2, \cdots, K^*(\lambda)\}$ such that $\text{sign}(v^*_{j-1}(\lambda)- v^*_{j}(\lambda)) = \text{sign}(v^*_{1}(\lambda)-y_{i})$, then the new restoration $\hat{u}$ satisfies $\hat u_{i-1}(\lambda) = u^*_{i}(\lambda)$ for all $i\geq i^{j+1,*}_1$.
    \label{thm:diff_bloque_head}
\end{thm}
To sum up, for updating $u^*(\lambda)$ to $\hat{u}(\lambda)$, only the first part and the last part of $w^m_i$ are changed. We introduce the following definitions:
\begin{definition} With $l = i^j_1$ and $p = i^{k+1}_1-1$ where $j$ is the last segment which satisfies $\text{sign}(v^*_{j-1}(\lambda)- v^*_{j}(\lambda))= \text{sign}(v^*_{K^*}(\lambda) - y_{i+m})$ and $k$ is the first segment which satisfies $\text{sign}(v^*_{k-1}(\lambda)- v^*_{k}(\lambda))= \text{sign}(v^*_{1}(\lambda)-y_{i})$,
    \begin{itemize}
        \item $(u^*_l, \cdots, u^*_m)$ is called \emph{non right-isolated sequence}.
        \item $(u^*_2, \cdots, u^*_p)$ is called \emph{non left-isolated sequence}.
        \item $(u^*_{p+1}, \cdots, u^*_{l-1})$ is called \emph{isolated sequence}.
    \end{itemize}
\label{def:non-iso}
\end{definition}

\subsection{Independence between segments}
For a given $\lambda$ (noted $\hat \lambda$), the cutting set $\mathcal{N}(\hat\lambda)$ is given by $\{\lambda^{\circ} > \hat \lambda\}$. In \cite{Liu}, we showed the points inside a segment of $u^*(\hat \lambda)$ are merged for $\lambda\leq \hat \lambda$, and the value of $\lambda$ provoking the merge is independent to the points outside the segment. By introducing the virtual segment (c.f. Definition \ref{def:virtual}), the estimation of $\Lambda^{\circ}$ can be broken into some sub-problems for each virtual segment, shown in proposition \ref{pro:online}. Combining with Theorem \ref{thm:diff_bloque} and \ref{thm:diff_bloque_head}, the elements of $\{\lambda^{\circ} \leq \hat \lambda\}$ inside the isolated sequence are not influenced by the sliding movement. 

\begin{definition}
  Let $\hat\lambda$  and $\epsilon_\lambda >0$, we have $v^*(\hat\lambda) = \{v_1^*(\hat\lambda), \cdots, v_l^*(\hat\lambda)\}$, $\mathcal{N}^*(\hat\lambda) = \{\mathcal{N}_1, \cdots, \mathcal{N}_l\}$, $l=K(\hat\lambda)$ and $s_i= \text{sign}(v_{i+1}^*(\hat\lambda)-v_i^*(\hat\lambda))$. For each segment $\{y_{\mathcal{N}_j},\tau_{\mathcal{N}_j}\}$ with $j =1, \cdots,l$, let $c_i = \frac{\hat\lambda+\epsilon_\lambda}{2s_i}$, we introduce the \emph{virtual segment} $\{y^+_{\mathcal{N}_j},\tau^+_{\mathcal{N}_j}\}$ where:
  \begin{itemize}
      \item  $y^+_{\mathcal{N}_1}=  \{y_{\mathcal{N}_1}, v_1^*(\hat\lambda) + c_1\}$, $y^+_{\mathcal{N}_i} =  \{v_i^*(\hat\lambda) - c_{i-1}, y_{\mathcal{N}_i},  v_i^*(\hat\lambda) + c_i\}$ for $ i = 2,..., l-1$ and $y^+_{\mathcal{N}_l} =\{v_{l}^*(\hat\lambda) - c_{i-1}, y_{\mathcal{N}_{l}}\} $.
      \item $\tau^+_{\mathcal{N}_1} = \{\tau_{\mathcal{N}_1}, 1\}$, $\tau^+_{\mathcal{N}_i} = \{1, \tau_{\mathcal{N}_i},  1\}$ for $i = 2,..., l-1$ and $\tau^+_{\mathcal{N}_l} = \{1, \tau_{\mathcal{N}_{l}}\}$.
  \end{itemize}

\label{def:virtual}
\end{definition}

\begin{pro}
Let $\Lambda^{\circ}$ the estimation with all the samples $\{y_i, t_i\}_{1,\cdots,n}$, $\Lambda^\circ_i= \{\lambda_{i,1}^\circ, \cdots\}$ the estimation with $i^{th}$ virtual segment $(y ^+_{\mathcal{N}_i}, \tau_{\mathcal{N}_i}^+)$, $\Lambda^{*}_1 = \{\lambda_{i,1}^\circ, \cdots, \lambda_{i,n_i-1}^\circ\}$ and $\Lambda^{*}_i = \{\lambda_{i,2}^\circ, \cdots, \lambda_{i,n_i}^\circ\}$ for $i = 2,..., l$,
we have $\cup_{i=1,\cdots, l} \Lambda^{*}_i= \{\lambda |\lambda \leq \hat\lambda \cap \lambda \in \Lambda^{\circ}\}$.
\label{pro:online}
\end{pro}

\subsection{Proposition of algorithms}
Let $\Delta g^{\circ} = (\Delta g(\lambda^{\circ}_1), \Delta g(\lambda^{\circ}_2), \cdots, \Delta g(\lambda^{\circ}_{n-1}))$, we note $\Lambda^{\circ,i}$ and $\Delta g^{\circ,i}$, two vectors of size $m-1$, the result of $w^m_i$. The results after sliding ($\Lambda^{\circ,i+1}$ and $\Delta g^{\circ,i+1}$) based on $w^m_{i+1}$ can be obtained by updating $\Lambda^{\circ,i}$ and $\Delta g^{\circ,i}$. In this section, we will propose an adaptation of the online algorithm proposed in \cite{Liu} to the sliding windows. 

We will only talk about the online estimation of $\Lambda^{\circ, i+1}$ from $\Lambda^{\circ, i}$ in detail. By following, we note an application of Algorithm 1 in \cite{Liu} to a given sequence of $\{y\}$ and $\{\tau\}$ as $\Lambda^\circ = \text{DP-TV}(\{y\}, \{\tau\})$.


After choosing $\hat\lambda \geq 0$, called the \emph{cutting point}, the restoration for $w^m_i$ is $u^*(\hat \lambda)$. We note $p$ and $j^*_p$ respectively the last point and the last segment of the non left-isolated sequence of $u^*(\hat \lambda)$, $l$ and $j^*_l$ respectively the first point and the first segment of the non right-isolated sequence of $u^*(\hat \lambda)$. $\Lambda^{\circ,i+1}$ can be splitted into four parts: (1) Some elements of $\{\lambda^{\circ,i+1} > \hat \lambda\}$ are changed following Proposition \ref{pro:online}; (2) The non left-isolated sequence $\{\lambda^{\circ,i}_j \leq\hat{ \lambda}\}$ for $j\leq p$ is influenced by the removal of the first point $y_{i}$; (3) The non right-isolated sequence $\{\lambda^{\circ,i}_j \leq\hat{ \lambda}\}$ for $j\geq l$ is influenced by the new point $y_{i+m}$; (4) All $\lambda^{\circ,i}_j \leq\hat{ \lambda}$ remains the same for $p<j<l$.

We treat at first the unchanged part of $\Lambda^{\circ,i+1}$ : due to the sliding, we have $\lambda^{\circ,i+1}_{\{\lambda^{\circ,i+1}_{p, \cdots, l-2}  \leq \hat{\lambda} \}} = \lambda^{\circ,i}_{\{\lambda^{\circ,n}_{p+1, \cdots, l-1}  \leq \hat{\lambda}\}}$.


For the non right-isolated sequence, let $\epsilon_\lambda >0$, $\Lambda^a = \lambda^{\circ,i+1}_{\{\lambda^{\circ,i+1}_{l-1, \cdots, m}  \leq \hat{\lambda} \}}$ can be estimated by $\text{DP-TV}(y^+_a,\tau^+_a )$ with the virtual segment:
\begin{itemize}
    \item $y^+_a = \{v^*_{j^*_l}(\hat\lambda ) - \frac{\hat\lambda+ \epsilon_\lambda}{2\text{sign}(v_{j^*_l}-v_{j^*_l-1})}, y_{\{l+i-1,\cdots,m+i\}}\}$
    \item$\tau^+_a = \{1, \tau_{\{l+i-1,\cdots,m+i\}}\}$
\end{itemize}

For the non left-isolated sequence, $\Lambda^b = \lambda^{\circ,i+1}_{\{\lambda^{\circ,i+1}_{1, \cdots, p-1}  \leq \hat{\lambda} \}}$ can be estimated by $\text{DP-TV}(y^+_b,\tau^+_b )$ with the virtual segment:
\begin{itemize}
    \item $y^+_b = \{y_{\{i+1,\cdots,i+p-1\}}, v^*_{j^*_p}(\hat\lambda) + \frac{\hat\lambda + \epsilon_\lambda}{2\text{sign}(v_{j^*_p+1}-v_{j^*_p})}\}$
    \item$\tau^+_b = \{\tau_{\{i+1,\cdots,i+p-1\}},1\}$
\end{itemize}

The isolated and non-isolated sequences can be assembled in $\Lambda^c = \{\lambda^c_1,\cdots, \lambda^c_m\}$ with $\lambda^{c}_k=\lambda^a_{k-l+3}$ for $k\geq l-1$, $\lambda^{c}_k=\lambda^b_{k}$ for $k\leq p-1$ and  $\lambda^{c}_k=\lambda^{\circ, n}_{k-1}$ for $p<k-1<l$.


It remains $\{\lambda^{\circ,i+1}> \hat \lambda \}$. Let $d =  \{\lambda^{\circ,i+1} > \hat{\lambda}\} = \{\lambda^{c}_k> \hat{\lambda}\} $ containing indeed all the last points of $\hat{u}(\hat\lambda)$'s segments, we can get $\Lambda^d=\lambda^{\circ,i+1}_d = \text{DP-TV}(\hat{v}(\hat{\lambda}), \hat{\mathcal{T}}(\hat{\lambda}))$. 

Finally, $\Lambda^{\circ, i+1}$ can be assembled in the following way:
\begin{equation}
\lambda^{\circ, i+1}_k=\begin{cases}
\lambda^c_{k} , & \text{if } \lambda^{d}_k\leq \hat{\lambda}.\\
\lambda^d_{\alpha(i)} & \text{if } \lambda^{d}_k >\hat{\lambda}.

\end{cases}
\label{equ:Lambda_n+1}
\end{equation}
with $\alpha: \mathbb{Z} \to \mathbb{Z}$ giving the index of $i$ in the vector $d$. 

\begin{algorithm}
\caption{Adaptation of online implementation to sliding window}
\begin{algorithmic} 
\Require $(y_i, \cdots, y_{i+m-1}), (\tau_i,\cdots, \tau_{i+m-1}), \ (y_{i+m}, \tau_{i+m})$
\Require $\Lambda^{\circ,i}, \ \Delta g^{\circ,i},\ \hat{\lambda}$
\State Find non right-isolated sequence $(l,\cdots,m)$ of $u^*(\hat\lambda)$
\State Find non left-isolated sequence $(2,\cdots,p)$ of $u^*(\hat\lambda)$
\If{$p<l-2$} 
\State ($\Lambda^a,\Delta g^a) = \text{DP-TV}(y^+_a, \tau^+_a)$
\State ($\Lambda^b, \Delta g^b) = \text{DP-TV}(y^+_b, \tau^+_b)$
\State $\Lambda^{\circ,i+1}= \{\Lambda^b_{\{1, \cdots,p-1\}}, \Lambda^{\circ,i}_{\{p+1,\cdots,l-1\}}, \Lambda^a_{\{2, \cdots,m-l+2\}}\}$
\State  $\Delta g^{\circ,i+1}= \{\Delta g^b_{\{1, \cdots,p-1\}}, \Delta g^{\circ,i}_{\{p+1,\cdots,l-1\}},$
\State $\qquad\qquad \quad \Delta g^a_{\{2, \cdots,m-l+2\}}\}$
\State $d =  \{\lambda^{\circ,n+1} > \hat{\lambda} \}$

\State $(\lambda^{\circ,i+1}_d, \Delta g^{\circ,i+1}_d)  = \text{DP-TV}(\hat{u}(\hat{\lambda}), \hat{\mathcal{T}}(\hat{\lambda}))$
\Else \Comment{Offline approach}
\State $(\lambda^{\circ,i+1}_d, \mathcal{T}^{\circ,i+1}_d)  = \text{DP-TV}(\{y_i,\tau_i\}_{i+1,\cdots, i+m})$
\EndIf
\Return $\Lambda^{\circ,i+1}$ and $\Delta g^{\circ,i+1}$
\end{algorithmic}
\label{algo:online}
\end{algorithm}

The algorithm adapted for the sliding window is gathered in Algorithm \ref{algo:online}. With a nice choice of $\hat \lambda$, only a small part ($\Lambda^a, \Lambda^b, \Lambda^d$) of the window $w^m_{i+1}$ needs to be updated from $w^m_{i}$. The overall complexity for estimating the restoration in a window of size $m$ is in $O(m)$.

\section{Performance analysis}
\label{sec:app}
\subsection{Application: noise variance monitoring}
We propose a method to detect the shift of noise's variance based on our restoration method. For an observed signal $y$ of size $n$, we take a sliding window of size $m$ with $m<n$. For each window $w_i^m = [i, i+m-1]$, we apply our denoising algorithm with the automatic choice of $\lambda$ on the sequence $(y_i, \cdots, y_{i+m-1})$ and $(\tau_i, \cdots, \tau_{i+m-1})$ for estimating the restored signal $u^*(\lambda_{ours}) = (u^*_1, \cdots, u^*_m )$. The windowed restoration residual for $j = 1, \cdots, m$ is given by $r_j = y_{j+i-1} - u^*_j(\lambda_{ours})$. The variance of the residual inside $w_i^m$ can be estimated by: 
\begin{equation}
    (\sigma_i^{m*})^2 = \frac{1}{m-1} \sum_{j = 1}^{m} (r_j - \overline{r}_i)^2
    \label{equ:sigma_res}
\end{equation}
with the mean value of residual $\overline{r}_i = \frac{1}{m} \sum_{j = 1}^{m} r_j$.

For the simulation, the realisation of noise is $\hat \epsilon_j = y_j - u_{net,j}$ with $u_{net}$ the original signal and $j = 1,\cdots, n$. The estimation of noise variance inside $w_i^m$ is given by $(\hat\sigma_i^m)^2 = \frac{1}{m-1} \sum_{j = i}^{i+m-1} (\hat \epsilon_j - \overline{\hat \epsilon}_i)^2$ with $\overline{\hat \epsilon}_i = \frac{1}{m} \sum_{j = i}^{i+m-1} \hat \epsilon_j$.

Since $u^*(\lambda_{ours})$ is a good restoration of $y$, $u^*(\lambda_{ours})$ is similar to $u_{net}$, which means that $\sigma_i^{m*}$ is close to $\hat\sigma_i^m$ inside each window $w^m_i$. The noise variance $\sigma_i^{m*}$ is not available for the real data, so we propose to detect the variance shift in using the residual standard deviation vector $\sigma^{m*} = (\sigma^{m*}_1,\cdots, \sigma^{m*}_{n-m+1})$ with $\sigma^{m*}_{i}$ obtained by (\ref{equ:sigma_res}) inside $w^m_i$.

For estimating the noise variance in a window of size $m$, the complexity is in $O(m)$ with the online implementation of sliding windows, and the overall complexity for monitoring the variance of a signal of size $n$ is in $O((n-m)m)$.

\subsection{Results}
For the variance monitoring, the method needs to estimate accurately the noise variance in the ideal case or capture the variation of noise variance in a more realistic case. We use the following criteria to evaluate the performance of variance monitoring:
\begin{itemize}
    \item Average bias: $\overline{\text{bias}} = \frac{1}{n-m+1}\sum_{i=1}^{n-m+1}(\hat\sigma^m_i - \sigma^{m*}_i)$.
\item Ratio of $\hat\sigma^{m}$ variation explained: 
\begin{equation}
    \text{RVE} = 1-\frac{\sum_{i=1}^{n-m+1}(\hat\sigma^m_i - \overline{\text{bias}} -\sigma^{m*}_i)^2}{\sum_{i=1}^{n-m+1}(\hat\sigma^{m}_i - \overline{\hat\sigma^{m}})^2}
\end{equation}
where $\overline{\hat\sigma^{m}} = \frac{1}{n-m+1}\sum_{i=1}^{n-m+1}(\hat\sigma^m_i)$.
It is indeed the $R^2$ score between $\sigma^{m*}$ and $\hat\sigma^m$ after adjusting those two items to the same mean value. The range of $\text{RVE}$ is between $0$ and $1$, and $\text{RVE}=1$ indicates our estimation explains perfectly the variation of $\sigma^{m*}$.
\end{itemize}

We compare our method with Median Absolute Deviation (\emph{MAD}) following the proposition of  \cite{donoho1994ideal} and \cite{sardy2016threshold} $\sigma = 1.4826\ \text{median}(By - \text{median}(By))$
with $By = \sqrt{2}(y_2-y_1,\cdots, y_n - y_{n-1})$. 
\begin{figure}[!htb]
\centering
\subfloat{
\includegraphics[width=\textwidth/2]{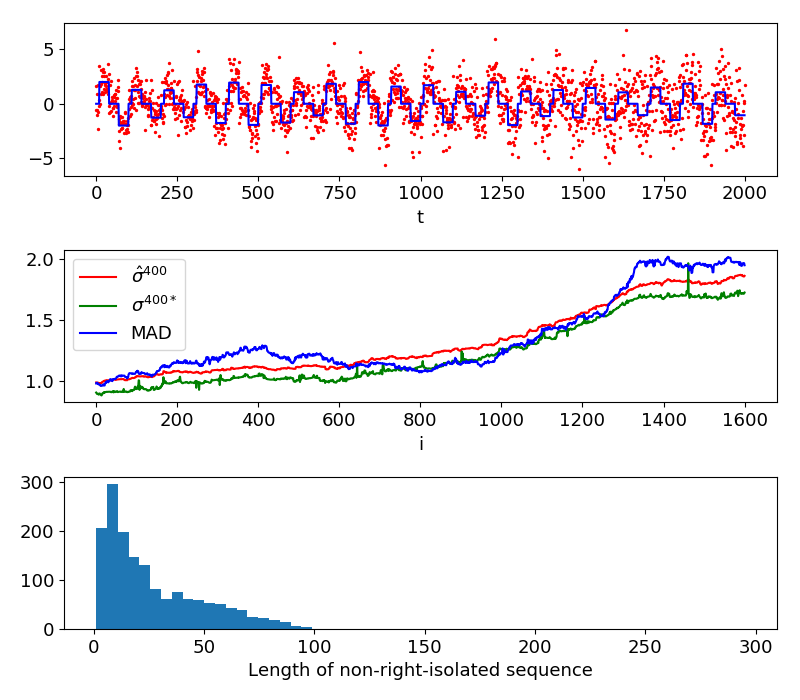}
}
\caption{Up: example of simulated signal: $y = u_{net} + \epsilon$ with $\epsilon \sim \mathcal{N}(0, \sigma_a^2(t))$. Colors : $y$ (red) and $u_{net}$ (blue). Middle: the standard deviation of the noise realisation $\epsilon = y - u_{net}$ (red line), that estimated by our method (green line) and by MAD (blue line) inside every $w^{400}_i$. Down : histogram of non-right-isolated sequence's length inside each sliding window }
\label{fig:exe}
\end{figure}

At first, we will fix the window length $m = 400$. Our proposition requires a parameter $q$ which stands for the length of the approximation of derivative, see p13 in \cite{Liu} for more details. We apply our method to a simulated piece-wise constant signal with the parameter $\log_{10}(q) =1$. 
An example of $y = u_{net} + \epsilon$ with $\epsilon \sim \mathcal{N}(0, \sigma_a^2(t))$ and the noise variance estimations for 400-point windows $w^{400}_i$ and $i = 1, \cdots, 1601$ are shown in Figure \ref{fig:exe}. Both MAD and our method propose an estimation similar to the variance of the noise realisation for every sliding windows. In this example, our method underestimates the variance, but the variation of our estimation inside each sliding window follows well that of $\hat \sigma^{400}$. The estimation of MAD propose a precise estimation, but does not follow the variation of $\hat \sigma^{400}$. The lengths of non-right-isolated sequence of each window (shown in Figure \ref{fig:exe}) are all smaller than the window size $m=400$, which means the influence of new sample remains inside each window and validates our assumption about the local influence.

We tested the performance of MAD and our approach under different noise hypothesis:
\begin{enumerate}
    \item  $\mathcal{N}(0, \sigma_a^2(t))$ with $\sigma_a(t) = 1 + 0.0005t$.
    \item $\mathcal{N}(0, \sigma_b^2(t))$ with $\sigma_b(t) = 1$ for $t\leq 1000$ and $\sigma_b(t) = 1 + 0.001 t$ for $t>1000$.
    \item $\mathcal{U}(-d(t), d(t))$ with $d(t) = 1 + 0.0005t$.
    \item $\mathcal{N}(0, \sigma_a^2(t)) + \mathcal{U}(-1, 1)$.
\end{enumerate}

\begin{figure}[!htb]
\centering
\subfloat{
\includegraphics[width=\textwidth/2]{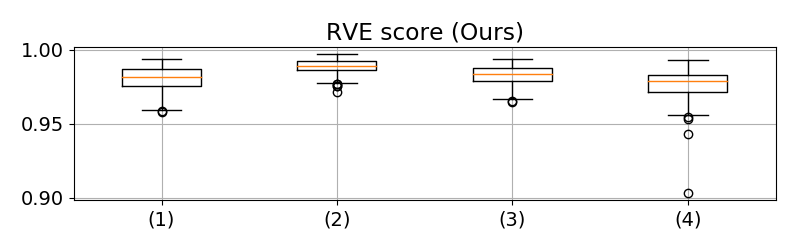}
\label{fig:r2_ours}
}
\vspace{-5pt}
\subfloat{
\includegraphics[width=\textwidth/2]{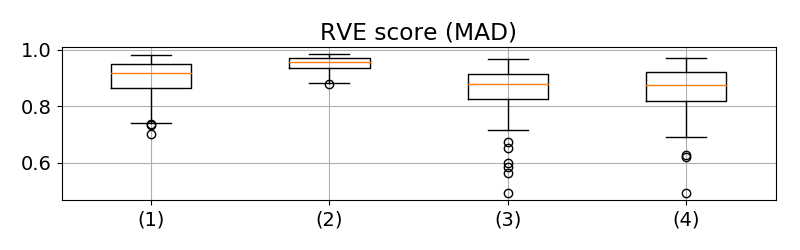}
\label{fig:r2_mad}
}
\vspace{-5pt}
\subfloat{
\includegraphics[width=\textwidth/2]{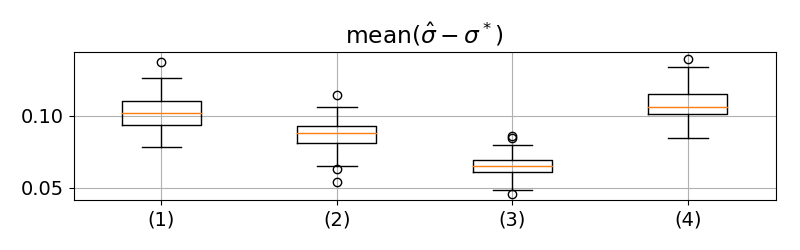}
\label{fig:e_ours}
}
\caption{Result of 100 simulations with different types of noise (1): $\mathcal{N}(0, \sigma_a^2(t))$, (2): $\mathcal{N}(0, \sigma_b^2(t))$, (3): $\mathcal{U}(-d(t), d(t))$ and (4): $\mathcal{N}(0, \sigma_a^2(t)) + \mathcal{U}(-1, 1)$. Up: $\text{RVE}$ Score of our method; Middle: $\text{RVE}$ score of MAD; Down: $\overline{\text{Bias}}$ of our method.}
\label{fig:signal}
\end{figure}
\begin{figure}[!htb]
  \centering
  \subfloat{
  \includegraphics[width=\textwidth/2]{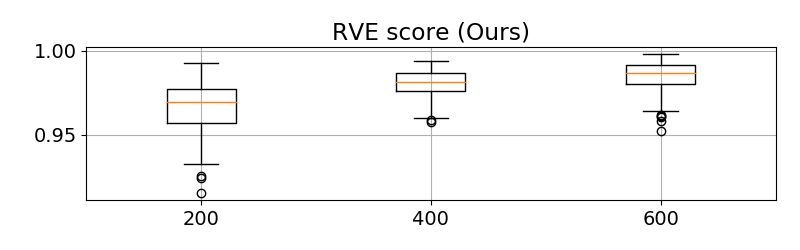}
  }
  \caption{$\text{RVE}$ score of 100 simulations with different window length (m). }
  \label{fig:point}
\end{figure}

We have done 100 simulations for each type of noise, and the results are gathered in Figure \ref{fig:signal}. For the most of simulations, our method has $\text{RVE}> 0.95$ and over-performs MAD. The high $\text{RVE}$ score indicates the variation of our proposition fits well that of the real values, which allows us to monitor the variation of noise variance for the real signal collected from plants. However, the result of $\overline{\text{Bias}}$ shows our method always underestimates the noise variance, and the bias depends on the type of noise, which means we can not estimate an universal offset for removing this bias in a general case.

The windows length $m$ plays an important role in the restoration and the variance monitoring, and needs to be chosen carefully. A short window can capture the local variation of the noise variance, but the new point (i.e $(i+m)^{th}$ point) may still have a strong influence for the pass (i.e. $i-1, i-2, \cdots$), which limits the restoration performance for the pass points. We test $m = 200, 400, 600$ with the noise $\epsilon \sim \mathcal{N}(0, \sigma^2_a(t))$, and the results are shown in Figure \ref{fig:point}. A longer window provides a better performance of monitoring. But the local variation of $\sigma^2$ will be neglected for a long window, and the stationarity hypothesis inside the window is not validated anymore. The choice of $m$ is out of the scope of this article, and it remains an open question.

To sum up, our method can not provide a precise estimation of the noise's variance values, but our estimation follows well that of the real value with an estimation error depending on the type of noise and the shape of original signal. 

One of the applications of our method is to detect the shift of noise variance. The residual variance may stay stable (i.e $\sigma^{m*}_i = \sigma$) during the correct functional period, and the noise variance shift can be detected by a prefixed threshold over $\sigma^{m*}_i$ comparing to the stable regime: e.g $\sigma^{m*}_i > 1.2\sigma$ for some successive sliding windows $w^m_i$.

         
 

\section{Conclusion}
In this article, we adapt the TV denoising method proposed in \cite{Liu} to sliding windows based on the local property of the TV-denoising signal. We believe our adaptive choice of $\lambda$ works also for the non-stationary signal. We applied the adaption to some signals with time varying noise variance. The simulated results shows that the variance of the restoration residuals follows well that of the noise, and this method can be used to monitor the ground noise variance in real time with a limited computation resource. However, the performance of these methods (both restoration and variance monitoring) depends on the choice of window length $m$ which is our on-going research interest. The application to the real data is also one of on-going works. 


\section*{Acknowledgment}
This work has been supported by the EIPHI Graduate School (ANR-17-EURE-0002).

\bibliographystyle{IEEEtran}
\bibliography{ecc2021}

\end{document}